# Diagnostics Using Nuclear Plant Cyber Attack Analysis Toolkit

Japan K. Patel[1], Athi Varuttamaseni[2], Robert W. Youngblood III[3], and John C. Lee[1]
[1]University of Michigan, Ann Arbor, MI; [2]Brookhaven National Laboratory, Upton, NY; [3]Idaho National Laboratory, Idaho Falls, ID

## INTRODUCTION

We introduce a Python interface for the GPWR Simulator [1] that allows users to define their cyber-attack scenarios, insert them into live plant simulation, collect realistic runtime data for analysis, and test diagnostics algorithms. The nuclear plant cyber-attack analysis toolkit (NPCAT) is designed to allow users to analyze plant responses to command-inject , denial-of-service, and man-in-the-middle type attacks. It comes with a built-in evaluation module comprising several diagnostics algorithms [2] including physics-based diagnostics with Kalman filtering, and data-driven diagnostics to detect anomalies. Our study focusses on modeling attacks on the steam generator (SG) level control system [3] using the GPWR Simulator. Our preliminary observations indicate that physics-based diagnostics with Kalman filtering is both robust and efficient. Additionally, support vector machines provide a good data-driven option for attack diagnostics. Considering recent advancements in quantum machine learning technology [4], an experimental quantum kernel [5] has been added to the framework for preliminary testing. First impressions suggest that while quantum kernels can be accurate, just like any other kernel, their applicability is highly problem/data dependent and can be prone to overfitting. To our knowledge, this is the first time a quantum algorithm has been explored for the purpose of detecting cyber-attacks on nuclear power plants.

## NUCLEAR PLANT CYBER ATTACK ANALYSIS TOOLKIT

NPCAT has been developed using Python 3.7 and leverages existing libraries like ScikitLearn [6], ScikitOptimize [7], and Qiskit [8], among others [3]. A JSON input file provides users with a convenient interface to customize plant operation states for cyber-attack analysis. The toolkit allows users to simulate attacks for several different operating conditions, power levels, and core states [2]. Transients like power escalation and de-escalation can be introduced seamlessly using turbine load follow maneuvers. Step and ramp type attack signals can be inserted along with synthetic noise to make attacks more realistic.

The GPWR source code is modified to make the workflow more convenient. Specifically, the SG level control code is altered to write {core thermal power, narrow range water level, feedwater flowrate, steam flowrate, valve position, SG power, secondary side inlet temperature, average reactor coolant system (RCS) temperature, SG pressure, global system pressure, SG saturation temperature, and simulation time} to text files. The controller source code has also been modified to plug in spoofing signals for sensors, controllers, and valves using an external script. Additionally, the source code is modified to make the three-loop plant mimic a two-loop design operating at a lower power.

We employ Latin hypercube sampling (LHS) [9] to generate the attack signals. NPCAT wraps around the GPWR database management system and performs user-defined simulations. The user input section of the interface obtains attack specifics, control system parameters to be spoofed, and how many samples to draw. This makes it possible to spoof anywhere from a single sensor to a set of all eight sensor and controller parameters associated with the SG level control system (SGLCS) [5]. Subsequently, the upper and lower limits of the relevant spoofing signals are defined. The interface, then, generates spoofing signals and stores these inputs in the form of a data-frame. NPCAT then initializes the GPWR run, inserts relevant spoofing signals into the simulation, and collects runtime data. The simulation time is monitored and if it exceeds a predefined limit or if the water level goes out of range, the simulation is terminated and new run with the next set of spoofing signals is started. Upon completion, trip times and respective spoofing signals are written to a csv file. This dataset is then used for analysis and relevant surfaces describing trip times can be plotted. The workflow is summarized in Figure 1.

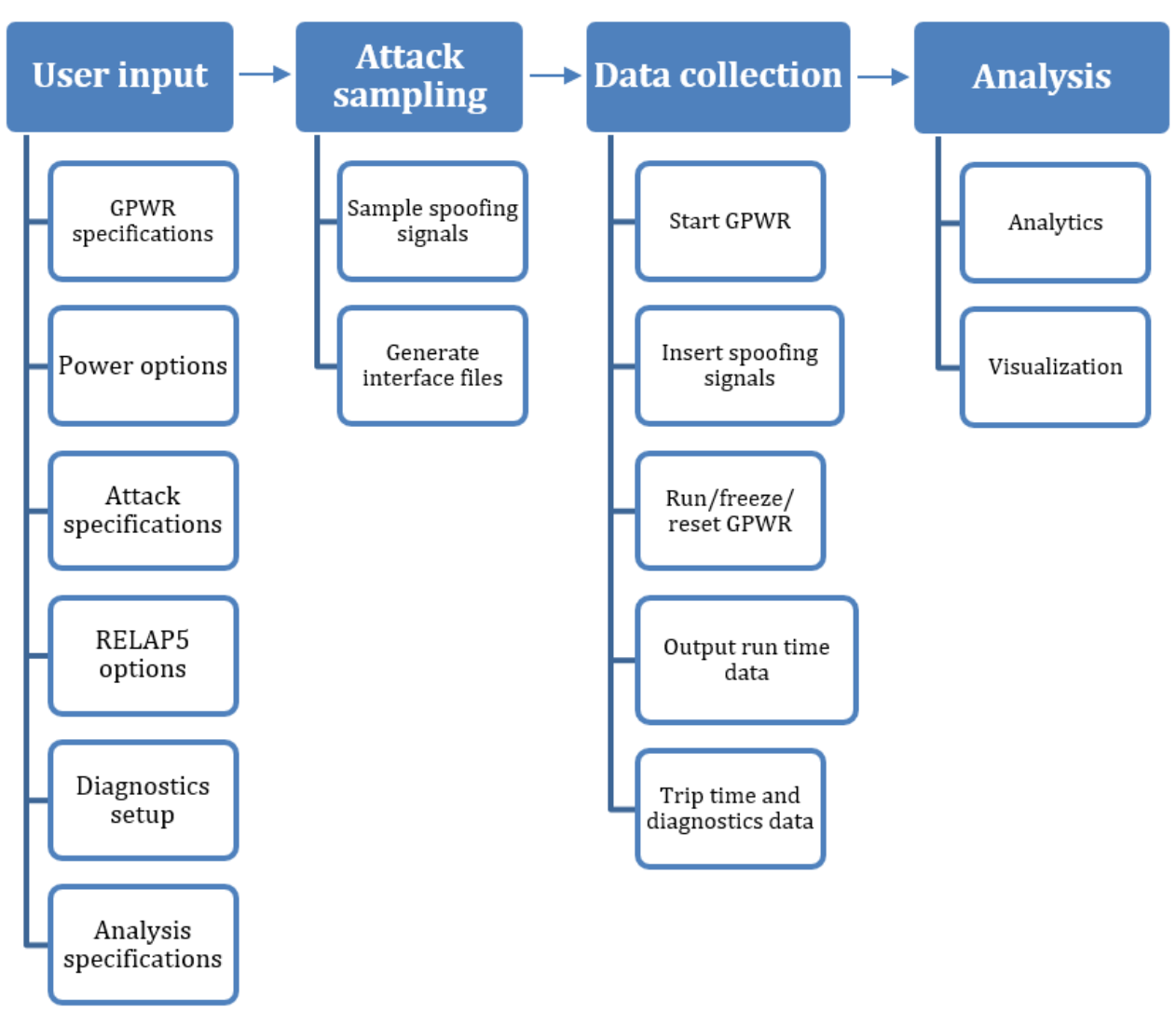

Figure 1. NPCAT pipeline.

## Evaluation Module

The components of the SGLCS can be susceptible to different types of cyber-attacks. These attacks can be classified as: (1) command inject (CI), (2) man in the middle (MiM), and (3) denial of service (DoS). Command inject attack occurs when a command is maliciously sent to one or more actuator components. An example of this is when a "valve open" or "valve close" command is sent to a valve positioner. Measurement and polling injection attacks occur when signals from the sensors or poll-generating equipment are spoofed. This is useful to an attacker in a "denial of information" setting where the attacker wishes to prevent the operators from obtaining true information. When used in conjunction with the command injection attack, the attacker can prevent the operators or control/protection systems from receiving the true status of plant equipment. This combination of attacks is also referred to as "man in the middle" attack. A denial-of-service attack occurs when the attacker prevents valid signals from reaching the intended destination. This can cause sensor readings to be delayed or actuation signals to reach the actuator late. The toolkit is designed so that each of the components in the model responds appropriately to these attack classes.

The evaluation module utilizes several diagnostics algorithms including physics-based diagnostics (PBD), online sensor validation (OSV), noise profiling (NP), and data-driven diagnostics (DDD). While the mainstay of DDD is support vector machine with a radial basis function kernel, we have added an experimental quantum kernel to the framework for preliminary testing. Quantum bits (qubits) exist in both zero and one states simultaneously while classical bits do not. In an ideal scenario, this provides qubits with the advantage of being able to represent and perform exponentially larger computations compared to classical bits [4]. To some extent, using qubits allow us to shift the burden of computing from hardware to algorithms, while being able to represent complex nonlinear surfaces relatively well. Quantum algorithms provide novel ways to transform data for classification using support vector machines. This has motivated our interest in exploration of quantum kernels.

## Physics-based reduced order model

A reduced order model is derived using mass and energy conservation. A simple boiling channel model is used to represent the dynamics of a U-tube SG and simulate SG-related cyber-attack scenarios. Our primary purpose is to obtain the water level and feedwater flowrate via a first-principles energy balance equation for the boiling channel, which is coupled to the primary coolant channel by representing the heat flux via the average temperature of the primary coolant flow. The vertical channel approximately represents the combination of co- and counter-current heat transfer processes inherent in a U-tube SG [10,11].

For a SG of length $H$, cross-sectional area $A$, and feedwater mass flowrate $W_s$, flow is divided into single- and two-phase regions separated by the water level at $z_0$. The heat flux $q_s(z)$ for the single-phase region is represented in terms of the fluid temperature $T_s(z)$, primary coolant temperature $T_p$, heat capacity $C_s$, effective heat transfer coefficient $U_1$ and wetted perimeter $M$:

$$W_s C_s \frac{dT_s(z)}{dz} = MU_1\left[T_p - T_s(z)\right], \qquad (1)$$

which is readily solved for $T_s(z)$ in terms of the feedwater inlet temperature $T_{s,in}$

$$T_s(z) = T_p + \left(T_{s,in} - T_p\right) \exp\left(-\frac{MU_1}{W_s C_s} z\right), \qquad (2)$$

along with an explicit expression for the water level with the saturated steam temperature $T_{sat}$

$$z_0 = -\frac{W_s C_s}{MU_1} \ln\left(\frac{T_p - T_{sat}}{T_p - T_{s,in}}\right). \qquad (3)$$

The parameters $U_1$ and $M$ are obtained via regression [10]. The total heat transfer rate into the SG is equal to the total power produced in the core, with the SG exit quality $x_e$ and latent heat of vaporization $h_{fg}$:

$$P_{SG} = W_s\left[C_s\left(T_{sat} - T_{s,in}\right) + x_e h_{fg}\right]. \qquad (4)$$

## Kalman filter (KF)

To obtain an optimal system estimate for the feedwater flowrate, we have used KF [11]. We treat data from the GPWR simulator as the observation, subject to statistical fluctuations, and the reduced order model above as the simulation model subject to uncertainty. We represent system state $x_l$ via state transition matrix $\Phi$ with variance $Q$ and the observation $y_l$ via measurement matrix $M$ with variance $R$ at timestep $l$ according to

$$\begin{aligned} x_l &= \Phi x_{l-1} + w_l, < w_l^T w_l > = Q, \\ y_l &= M x_l + v_l, M = I, < v_l^T v_l > = R. \end{aligned} \qquad (5)$$

Taking the noise from both the measurements and the simulator into consideration, example case studies have been presented in [2,10]. Absolute errors in feedwater flowrate are used to flag potential intrusions with a tolerance of 1%.

## Data-driven diagnostics

Support vector machine (SVM) is a supervised learning algorithm that can classify data by finding a separator surface [6]. This surface is chosen such that it represents the largest separation margin between the different classes of data. There are several advantages to these methods which include effectiveness in high dimensional spaces, memory efficiency, and versatility. Support vector machines can accept different kernel functions for decision-making. We consider two kernels: radial basis and quantum. While the radial basis kernel is well established and leverages classical computers, the quantum kernel is a more recent development and promises to leverage quantum or hybrid quantum computing architectures in the future.

**Radial basis kernel**

The surface separating the data into normal and anomalous classes is generated using the radial basis function (RBF) kernel with the scikit-learn library. With training parameter $\gamma$, the radial basis kernel $K_{rbf}(x,x')$ is represented according to:

$$K_{rbf}(x,x') = e^{-\gamma\left\|x-x'\right\|^2}. \qquad (6)$$

The $\gamma$ determines the influence of each sample [6]. For this report, our feature space consists of normalized power, feedwater flowrate, steam flowrate, and the SG water level. The decision regarding whether the plant operating state is anomalous or not is determined by the classifier.

**Quantum kernel**

Quantum classification is characterized by interactions like superposition, interference, and entanglement followed by measurement of quantum states of qubits interacting within a quantum circuit [4]. In other words, we start with classical feature data, then map it to quantum states that interact via predefined circuits. Subsequently, we measure output states to glean information on which class the input data belongs to.

While a detailed exposition on quantum machine learning is presented in [4], we briefly review the use of quantum kernels in conjunction with support vector classification next. The general approach includes the following steps [4]:

1) Map classical features into quantum space using a state preparation circuit,
2) Quantum kernel estimation,
3) Prediction using classical SVM infrastructure.

A quantum classifier is implemented into NPCAT using IBM's qiskit library [8]. A second-order Pauli-Z evolution circuit [5] with three feature dimensions, two repeated circuits, and linear entanglement is employed for state preparation [4]. The quantum kernel $K_q(\boldsymbol{x},\boldsymbol{x}')$ is estimated according to the overlap of the two quantum states [4] :

$$K_q(\boldsymbol{x},\boldsymbol{x}') = |< \boldsymbol{0}^{\otimes n}|H^{\otimes n}\boldsymbol{U}_{\phi}(\boldsymbol{x})\boldsymbol{U}_{\phi}(\boldsymbol{x}')H^{\otimes n}|\boldsymbol{0}^{\otimes n} >|^2, \qquad (7)$$

where $\boldsymbol{x}$ and $\boldsymbol{x}'$ are feature vectors, $H^{\otimes n}$ represents application of Hadamard gates to $n$ qubits, $\boldsymbol{0}^{\otimes n}$ is the ground state, and feature map is a unitary operator [5] $\boldsymbol{U}_{\phi}(\boldsymbol{x})$ of depth $d = 2$:

$$\boldsymbol{U}_{\phi}(\boldsymbol{x}) = \left(e^{i\sum_{j,k}\phi_{\{j,k\}}(\boldsymbol{x})\boldsymbol{Z}_j\otimes\boldsymbol{Z}_k}e^{i\sum_j\phi_{\{j\}}(\boldsymbol{x})\boldsymbol{Z}_j}\right)^d, \qquad (8)$$

with data mapping function [5]

$$\phi_S(\boldsymbol{x}) \rightarrow \begin{cases} x_j & if\ S = \{j\} \\ (\pi - x_j)(\pi - x_k) & if\ S = \{j,k\} \end{cases} \qquad (9)$$

where $S$ defines connectivity between qubits with indices $j$ and $k$, and $\boldsymbol{Z}$ represents the Pauli Z transformation matrix. This kernel is passed on to classical support vector machine using the scikit-learn library [6] and predictions are made using a classical computer. We note that quantum machine learning algorithms have a potential advantage only if the kernel is hard to compute classically [4]. Moreover, quantum advantage is highly dependent on the feature maps used and the type of data being classified. In this paper, feature vectors include difference in water level, feedwater flowrate readings error, and flow mismatch. The circuit and features were determined by trial and error. Why a specific circuit would work or wouldn't is still an open question that we plan on exploring in the future.

**RESULTS**

With plant behavior and reactor trips due to spoofing addressed at length in [2, 3], we evaluated the performance of diagnostics algorithms using a small cohort of ten test problems. The assessment is based on observed accuracy and efficiency: does the method diagnose an attack accurately, and if so, how soon? We consider these criteria keeping in mind the utility of diagnostics approaches to reactor operators. We want diagnostics method to alert operators as quickly as possible so they can prepare to counter malicious attacks and take corrective measures.

With the first sample representing an unspoofed case, the remaining nine illustrate man in the middle attacks. Steam generator water level transmitter (LT) and feedwater flowrate transmitter (FT) were attacked where sensor readings were perturbed randomly within a range of roughly 10% above and below the normal operating conditions. The GPWR Simulator was run at full power under normal operating conditions for 70 s without entering any spoofing signals to get a baseline on false positives returned by the diagnostics algorithms. The specific set of attack signals has been presented in Table I.

Table I. Attack parameters.

| Case # | $t_{insertion}$ [s] | LT spoofing [%] | FT spoofing [lb/s] |
|---|---|---|---|
| 1 | 3.0 | 64.1 | 1327.5 |
| 2 | 3.0 | 62.9 | 1303.9 |
| 3 | 6.0 | 61.8 | 1280.3 |
| 4 | 6.0 | 60.7 | 1256.7 |
| 5 | 9.0 | 59.5 | 1233.1 |
| 6 | 6.0 | 54.4 | 1126.9 |
| 7 | 6.0 | 53.3 | 1103.3 |
| 8 | 3.0 | 52.1 | 1079.7 |
| 9 | 3.0 | 51.0 | 1056.1 |

Table II summarizes times taken by PBD and DDD to diagnose attacks: $t_{diag,\ KF}$, $t_{diag,\ SVM}$, and $t_{diag,\ qSVM}$, respectively. It also presents the time $t_{trip}$ it takes each set of spoofing signals to trip the reactor.

Table III. Diagnostics results.

| Case # | $t_{trip}$ [s] | $t_{diag, KF}$ [s] | $t_{diag, SVM}$ [s] | $t_{diag, qSVM}$ [s] |
|---|---|---|---|---|
| 1 | 91.2 | 8.5 | 17.7 | 8.5 |
| 2 | 106.3 | 8.7 | 23.8 | 11.5 |
| 3 | 128.1 | 11.5 | 35.1 | 8.7 |
| 4 | 158.5 | 11.5 | 46.1 | 8.7 |
| 5 | OT | 16.5 | 60.9 | FP |
| 6 | OT | 11.1 | 12.1 | 8.5 |
| 7 | OT | 11.1 | 12.1 | 8.4 |
| 8 | OT | 8.5 | 6.7 | 8.5 |
| 9 | OT | 8.4 | 6.7 | 8.4 |

The nine examples examined whether a man in the middle attack was diagnosed appropriately. The maximum runtime was set to 180 s and artificial noise was introduced only in the last sample. With OT representing cases that did not trip within three minutes of predefined runtime and FP representing false positive, Data-driven diagnostics was found to be kernel dependent as the SVM with RBF kernel did not return a false positive, but the quantum kernel did.

We observe that physics-based diagnostics with KF provided accurate diagnoses for all cases under consideration. This makes sense because as soon as the controller starts responding to malicious signals, it starts sending an abnormal amount of water to the steam generator. Therefore, even if the spoofing signals are masked well, mass and energy conservation will almost always reveal anomalies. This further suggests that having a small set of unhackable analog sensors in place that facilitate secure checks for energy and mass balance would enhance plant safety. The speed at which it diagnosed attacks remained roughly the same. For the first five cases, where feedwater flowrate spoofing signal was larger than the nominal value of 1181 lb/s, the detection time was shorter than SVM with RBF kernel. Concurrently, it was within a couple of seconds for the remaining cases. Observations suggest that for cases with smaller perturbations from nominal, SVM's separator surface is not as sharp. However, given that it diagnoses every case correctly, the model generalizes well and is robust against overfitting. SVM with the quantum kernel was observed to be quite efficient but returned false positives and therefore warrants further investigation with respect to robustness.

We also note that while NP consistently returned accurate diagnoses for most cases, it failed to detect an attack when the attack signal introduced artificial noise that was consistent with GPWR Simulator's inherent noise patterns. OSV was slower than other methods, in general, but certainly robust and always returns the correct diagnosis if all sensor channels were not attacked at the same time.

To summarize, we developed an application programming interface NPCAT for the GPWR Simulator and performed preliminary testing of the diagnostics algorithms implemented in NPCAT. The interface facilitates simulation of cyber-attacks on the simulator's steam generator level control system. Three kinds of attack have been implemented: command inject, denial of service, and man-in-the-middle. Additionally, several different diagnostics algorithms have been implemented for anomaly detection, and analyzed with respect to accuracy and efficiency. We observed that while physics-based diagnostics was the most robust and efficient, other methods including support vector machines with quantum kernels were also promising.

## ENDNOTES

Physics-based conservation laws and reduced order models derived from them are the most robust way to check for anomalies. Additionally, given our observations, further development and testing of the toolkit is imperative. Quantum kernels need further exploration, especially with respect to data curation, robustness, and general kernel setup.

## ACKNOWLEDGEMENT

This work was supported by the U.S. DOE Nuclear Energy University Program (NEUP) Grant DE-NE0008783, 18-15056. Additionally, we would like to thank Dr. Gokul Ravi for discussions on quantum computers and hardware.

## REFERENCES

[1] GSE Systems: Simulation + Solutions, 2020. Training and Simulation for Confident Operations. https://www.gses.com/systems.

[2] Lee, J. C., et al., 2022. Model-Based Diagnostics and Mitigation of Cyber Threats, Final Project Report. NEUP Project 18-15056.

[3] Patel J. K., Varuttamaseni, A. , Youngblood, R. W. , Wacker, S., Barbosa, R. P., Guo, J., and Lee, J. C., 2022. "Estimation of the time for steam generator trip due to cyber intrusions," *Ann. Nucl. Energy* 173, 109108.

[4] Petruccione, F. and Schuld, M., 2021. *Machine Learning with Quantum Computers*, Springer.

[5] Havlíček, V, et al., 2019. "Supervised learning with quantum-enhanced feature spaces," *Nature,* 567, 209.

[6] Pedregosa, F. et al., 2011. "Scikit-learn: machine learning in Python," *J. Machine Learn. Res*. 12, 2825

[7] Head, T. et al., 2020. "scikit-optimize/scikit-optimize(v0.8.1)," Zenodo. https://doi.org/10.5281/zenodo.4014775.

[8] Stein, M., 1987. "Large sample properties of simulations using Latin hypercube sampling," *Technometrics* 29 143.

[9] Qiskit contributors, 2023. "Qiskit: An Open-source Framework for Quantum Computing," 10.5281/zenodo.2573505.

[10] Guo, J., 2020. Model-Based Cyber Security Framework for Nuclear Power Plant. University of Michigan. Ph.D. Dissertation.

[11] Lee, J. C., 2020. *Nuclear Reactor Physics and Engineering*, Wiley.